# Traffic State Estimation in Congestion to Extend Applicability of DFOS

**Yoshiyuki Yajima[1]\*, Hemant Prasad[2], Daisuke Ikefuji[3],**
**Hitoshi Sakurai[4], Manabu Otani[5]**

1. NEC Corporation, yoshiyuki-yajima@nec.com
2. NEC Corporation
3. NEC Corporation
4. NEC Corporation
5. Central Nippon Expressway Company Limited

**Abstract**
This paper presents a traffic state estimation (TSE) method in congestion for distributed fiber-optic sensing (DFOS). DFOS detects vehicle driving vibrations along the optical fiber and obtains their trajectories in the spatiotemporal plane. From these trajectories, DFOS provides mean velocities for real-time spatially continuous traffic monitoring without dead zones. However, when vehicle vibration intensities are insufficiently low due to slow speed, trajectories cannot be obtained, leading to missing values in mean velocity data. It restricts DFOS applicability in severe congestion. Therefore, this paper proposes a missing value imputation method based on data assimilation. Our proposed method is validated on two expressways in Japan with the reference data. The results show that the mean absolute error (MAE) of the imputed mean velocities to the reference increases only by 1.5 km/h as compared with the MAE of non-missing values. This study enhances the wide-range applicability of DFOS in practical cases.

**1. Introduction**
Traffic monitoring is fundamental to visualize and recognize traffic flow as the first step for the intelligent transportation systems (ITS) toward future smart cities. Based on the traffic monitoring results, road operators can provide necessary information to drivers and control traffic flow to optimize it. Since the time and spatial scale of abnormal traffic events such as congestion are short-term and localized, real-time and wide-area monitoring for the entire road is essential. In terms of this point, the application of distributed fiber-optic sensing (DFOS) is promising [1].

Figure 1 shows the schematic view of traffic monitoring using DFOS. It detects and localizes vibrations along the optical fiber based on Rayleigh backscattered light. In traffic monitoring applications, the DFOS apparatus is connected to already installed fiber cables for telecommunications and detects vibrations generated by driving vehicles. By tracking vibration sources, vehicle trajectories can be obtained in DFOS raw data. Since the raw data severely suffers from optical noise and other environmental vibrations, the existing study [2] developed an image-processing-based traffic monitoring method for practical use. It derives mean velocity from slopes of extracted trajectories below the error of 10%. This technology has advantages in terms of real-time, installation-cost-free, power-efficient, wide-area without dead zones and high capability for traffic monitoring. Hence, DFOS has significant advantages compared with conventional sensors such as loop inductors, supersonic vehicle detectors, surveillance cameras, and probe vehicles. Toward ITS using DFOS, novel technologies for practical uses such as signal processing [3–5], a solution for the installation issue [6], and abnormal event detection [7] have been developed. It has also been reported that spatially continuous DFOS data helps to model traffic flow [8].

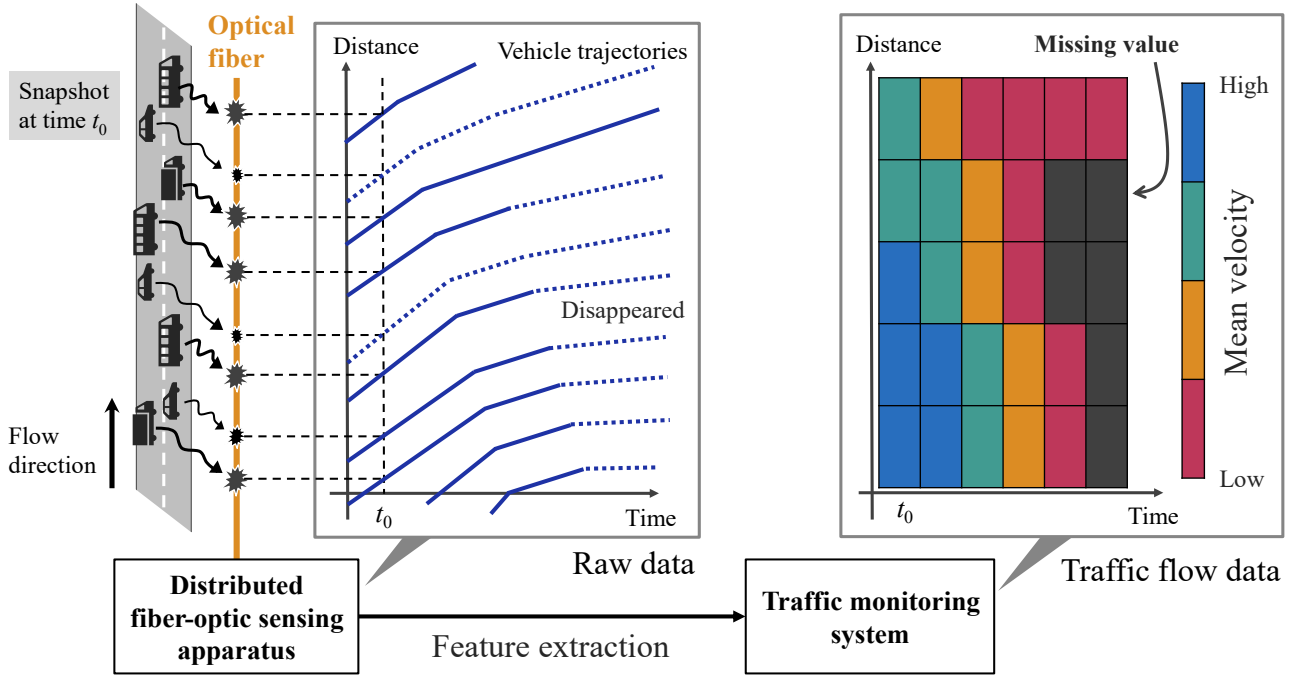

**Figure 1.** Schematic view of traffic monitoring using DFOS. When vibration is insufficiently weak to detect for the fiber cable due to slow vehicles in severe congestion, vehicle trajectories disappear. As a result, missing values occur in mean velocity data.

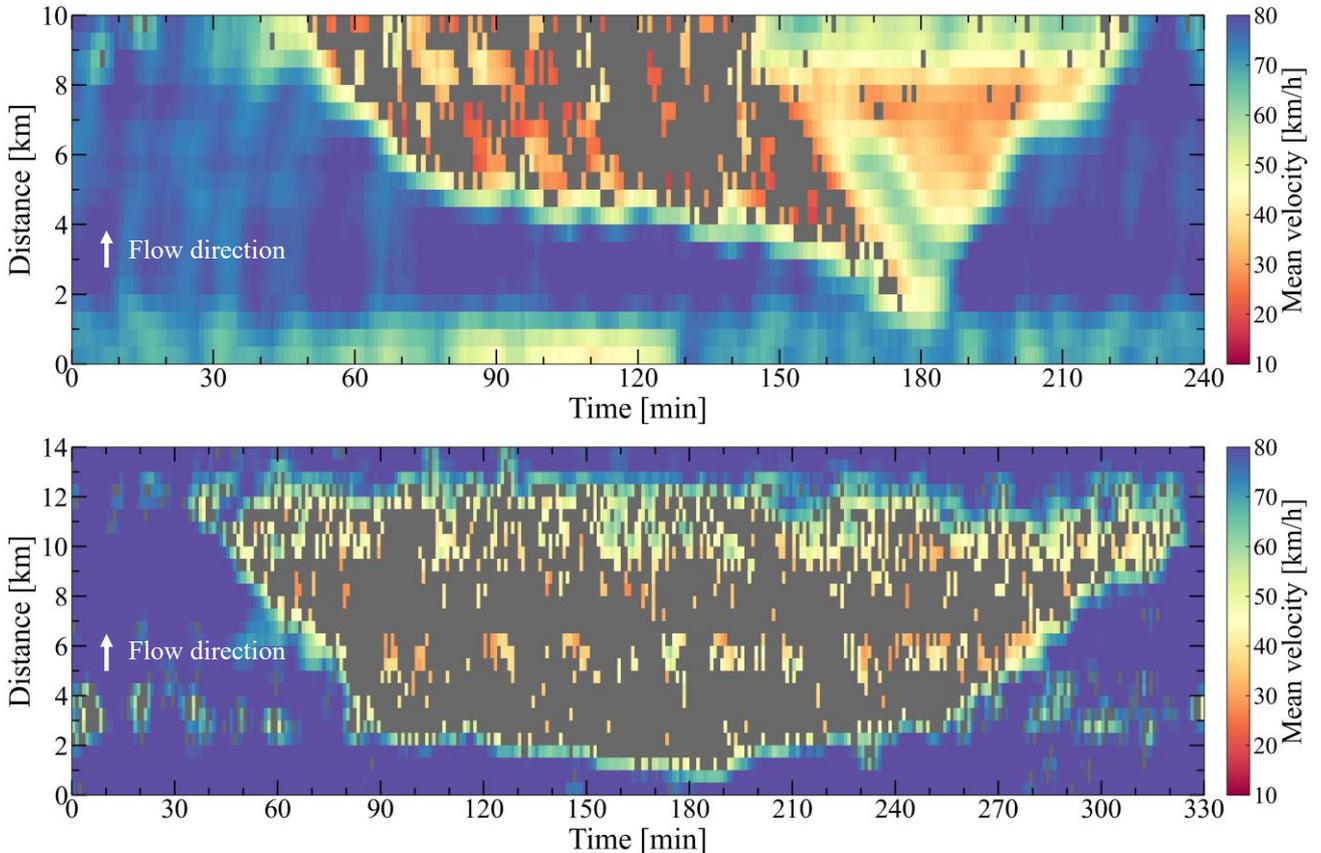

**Figure 2.** Observed mean velocity obtained from DFOS on (top) Ken-O Expressway and (bottom) Tomei Expressway in Japan. Each spatiotemporal patch is 500 m × 1 minute. Gray patches are missing values.

Although the application of DFOS is promising to realize ITS, there is a technical limit in severe congestion cases. Since DFOS detects driving vehicle vibrations, the vibration detection during severe congestion is sometimes challenging due to slow vehicles with low vibration intensity. As a result, vehicle trajectories disappear, and mean velocities cannot be derived, which results in missing values, as shown in Figure 1. Figure 2 shows the effect of slow vehicle vibrations resulting in missing values during actual congestion observed on two Japanese expressways. The occurrence of missing values is a serious issue in traffic monitoring and restricts the DFOS applicability despite its advantages. For instance, it is impossible to provide accurate traffic flow information during congestion, e.g., the travel time, without quantitative mean velocity data.

Conventional traffic state estimation (TSE) methods [9] are widely used for missing value imputation by reconstructing traffic flow data. TSE methods estimate traffic states outside the observed area from partially observed data. TSE is majorly classified into two categories: the data-driven and the model-driven approach. The former is based on an interpolation/regression model, time-series analysis, and deep learning [10,11]. Features such as common trends and patterns found in long-term data are extracted and the traffic state is estimated in dead zones of sensors. The latter is based on data assimilation that integrates observed traffic data and a theoretical model using the state-space model (e.g., Kalman filter) [12,13]. The combination of the two approaches has recently been proposed as physics-informed deep learning [14,15].

However, applying the existing TSE methods to DFOS data can be challenging. For instance, the data-driven TSE methods may not be the best appropriate approach because of the unavailability of sufficient DFOS data. Since DFOS-based traffic monitoring systems have emerged recently, observed traffic data from DFOS is limited. Due to the unavailability of long-term data, feature extractions in the data-driven TSE methods for DFOS data will be inaccurate. The application of existing model-driven TSE for DFOS data is also challenging. This is because the sensitivity of DFOS depends on the installation condition of the fiber cable. Therefore, the sensitivity changes from different road sections. As a result, all vehicle trajectories are not always observed, which may lead to underestimated mean flow rate and density in contrast to mean velocity [16]. Since existing model-driven TSE methods are based on the fundamental diagram with the three basic quantities (flow rate, velocity, and density), the existing model-driven TSE methods are not appropriate. This motivates us to develop a novel TSE method for DFOS systems using only mean velocity for the missing value imputation. This paper enables appropriate traffic monitoring and control using DFOS for the entire road with high visibility even in severe congestion.

## 2. Proposed Method for Missing Value Imputation

Figure 3 shows the outline of our proposed method. The brief overview is as follows. In step 1, simulation data sets that emulate traffic flow observed in the last 30 minutes are created with a theoretical model and multiple sets of traffic model parameters. In step 2, a prior probability distribution of mean velocity is set in each road segment. In addition, mean velocities measured by conventional traffic counters such as loop inductors are considered for road segments consisting of a traffic counter. In step 3, data assimilation using the particle filter is applied to the observed mean velocities and simulation data sets. In step 4, the optimal simulation scenario is identified from the posterior probability distributions of the model parameters. In step 5, missing values are replaced with the simulated mean velocities of the optimal scenario.

To create the simulation data sets in step 1, appropriate road parameters such as the road length, number of lanes, and bottleneck position are considered to model the respective road sections for the simulation. The traffic inflow rate and the initial speed of vehicles at the beginning of the road model are determined by the traffic counter data located at the corresponding position in the actual road. Based on these simulation settings, simulations adopted by multiple sets of model parameters are carried out. Changing the model parameter sets generates various traffic scenarios as shown in Figure 3 step 1.

This paper adopts the stochastic Nishinari-Fukui-Schadschneider (S-NFS) model [17]. It is a comprehensive cell-automata-based traffic model. The S-NFS model considers stochastic behaviors of vehicles: (i) involuntary deacceleration called the "random braking effect" with probability $p$, (ii) drivers' recognition delay or vehicle inertia called the "slow-to-start effect" with probability $q$, and (iii) drivers' anticipation called the "quick start effect" with probability $r$. In addition, the different random braking probability in the bottleneck $p^{\mathrm{BN}}$ introduced by our previous work [18] is considered to control the bottleneck strength. This paper refers to the model parameter set $\boldsymbol{\theta}$ as $\boldsymbol{\theta} = (p^{\mathrm{BN}}, p, q, r)$.

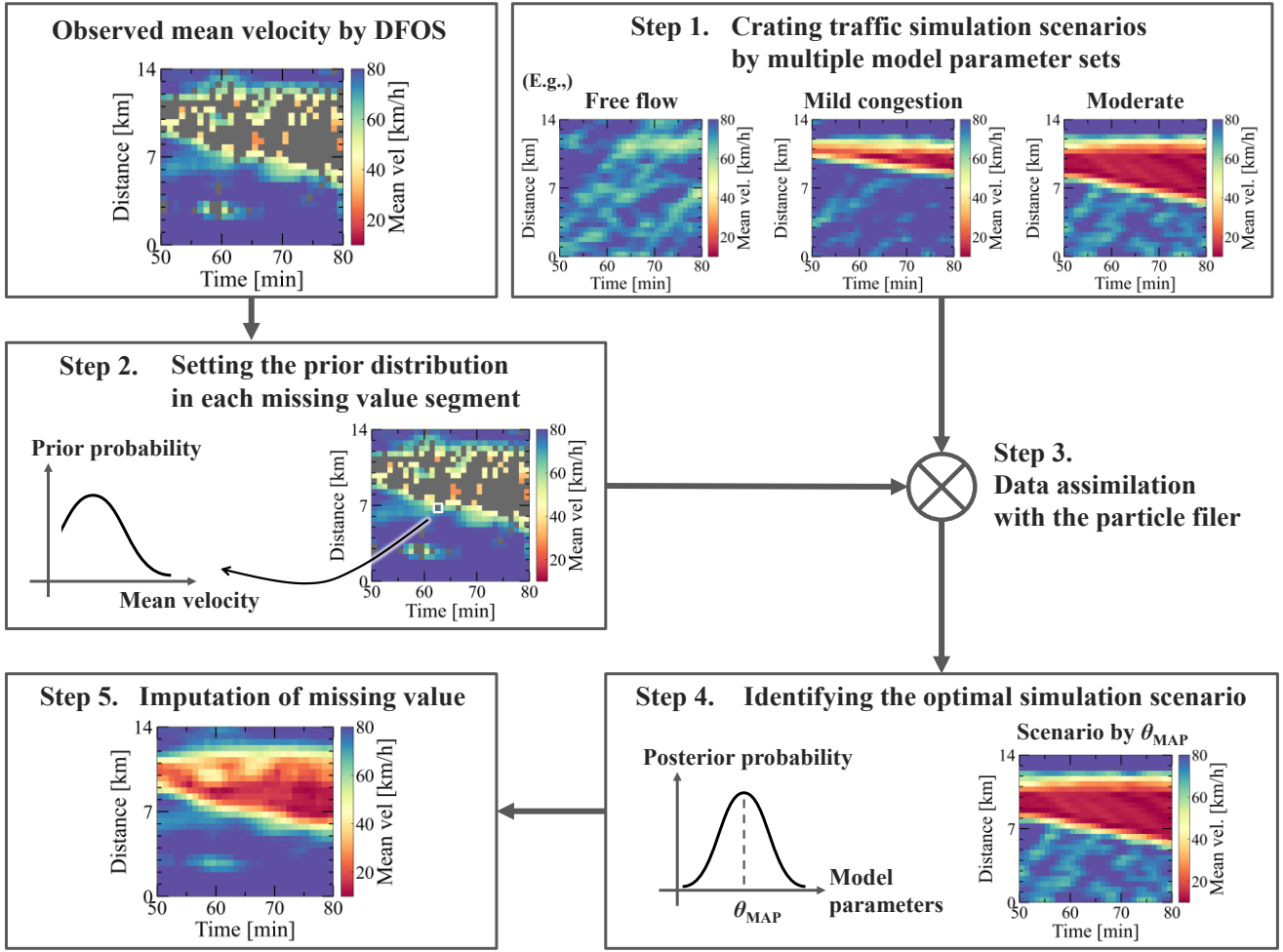

Figure 3. Outline of our proposed method for missing value imputation in DFOS mean velocity data.

To quantify mean velocities in each missing value segment, this paper introduces the prior probability distributions of mean velocity in step 2. The prior probability distributions take account of other sensor data (i.e., the sensor fusion), if it is available, and prior information expected from observed data. Here, based on the availability of traffic counters, missing value road segments are classified into two classes: segments consisting of a traffic counter are referred to as "class 1", and those without any traffic counters are referred to as "class 2". Figure 4 shows the prior probability distributions of mean velocity for class 1 and 2 road segments. The prior probability distribution $f(.)$ of the possible mean velocity $u$ for class 1 road segments is a truncated normal distribution considering the traffic counter data,

$$f(u) = \frac{1}{\sqrt{2\pi\sigma_a^2}Z} \exp\left[-\frac{(u-v_{\text{tc}})^2}{2\sigma_a^2}\right], \qquad u_{\min} \leq u \leq u_{\max}, \tag{1}$$

where $v_{\text{tc}}$ is the mean velocity observed by the traffic counter and $Z$ is the normalization factor. In class 2 segments, the prior distribution of missing values is defined as follows,

$$f(u) = \begin{cases} \dfrac{2}{u_{\max} + 20 - 2u_{\min}}, & u_{\min} \leq u \leq 20 \text{ km/h}, \\ \dfrac{2(u_{\max} - u)}{(u_{\max} - 20)(u_{\max} + 20 - 2u_{\min})}, & 20 \text{ km/h} < u \leq u_{\max}. \end{cases} \tag{2}$$

The threshold of 20 km/h is defined as the empirical tendency of DFOS data. Namely, the mean velocities obtained from DFOS without missing values are usually around 20 km/h at minimum although this lower limit depends on the installation environment of the fiber cable and road structures. Therefore, it is likely that the probability of $u \leq 20$ km/h is higher than the probability of $u > 20$ km/h. In addition, there is no traffic counter as a reference data in the class 2 road segments. This paper adopts the uniform distribution in $u_{\min} \leq u \leq 20$ km/h and monotonically decreasing linear function above 20 km/h.

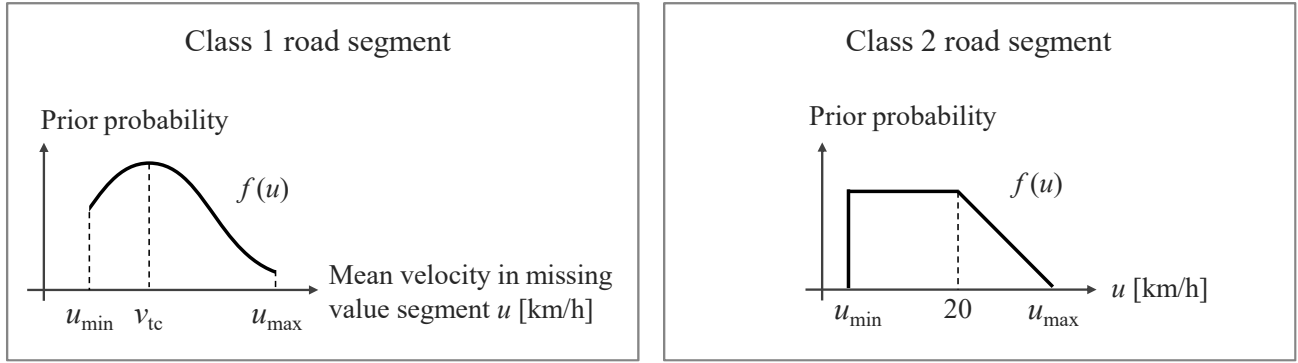

**Figure 4. Prior probability distributions for the class 1 and 2 road segments.**

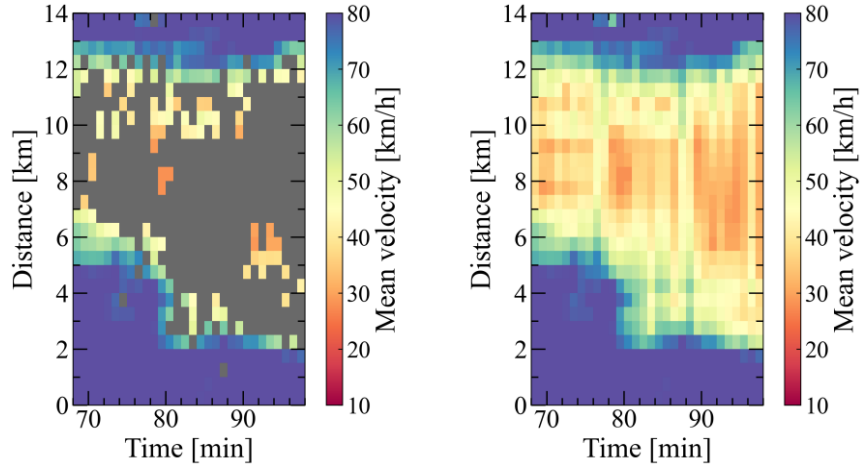

**Figure 5. (Left) Observed mean velocities in *t* = 68–98 minutes as shown in the bottom panel of Figure 2. (Right) The linear interpolation result along the time and spatial axis for the left panel.**

The lower and upper limit of the prior probability distributions $u_{\min}$ and $u_{\max}$ are defined as follows; $u_{\min} = 1$ km/h and $u_{\max}$ is the estimated mean velocity by linearly interpolating DFOS mean velocity data along the time and spatial axes. Figure 5 shows an example of linearly interpolated mean velocity data for missing values. The mean velocities in the missing value segments are expected to be lower than those in non-missing value segments, considering the reason for missing values due to low-speed vehicles. However, the estimated mean velocities from interpolation are similar to those of non-missing value segments as shown in Figure 5. The linear interpolation results in overestimated imputation, therefore, this paper considers the interpolated mean velocities to be the maximum limit for the prior probability distributions.

In step 3, the likelihood, or similarity between observed and simulation mean velocity in each scenario, at time *t*, is evaluated using the prior probability distributions of mean velocity in missing value segments. This likelihood is used in the particle filter process. According to our previous study [18], the likelihood of the simulation scenario by the *n*-th model parameter set $\boldsymbol{\theta}_n$ in the *m*-th segment at time *t*, $\mathcal{L}'_{m,n,t}(\boldsymbol{\theta}_n)$, in the road segments other than class 1 and 2, i.e., non-missing value segments is defined as

$$\mathcal{L}'_{m,n,t}(\boldsymbol{\theta}_n) = \mathcal{L}^p_{m,n,t}\left(E^p_{m,n,t}(\boldsymbol{\theta}_n)\right) \times \mathcal{L}^a_{m,n,t}\left(E^a_{m,n,t}(\boldsymbol{\theta}_n)\right), \qquad (3)$$

where $\mathcal{L}^p_{m,n,t}$ and $\mathcal{L}^a_{m,n,t}$ are a part of the likelihood functions defined by the following equations,

$$\mathcal{L}^p_{m,n,t}\left(E^p_{m,n,t}(\boldsymbol{\theta}_n)\right) = \frac{1}{\sqrt{2\pi\sigma_p^2}} \exp\left[-\frac{\{E^p_{m,n,t}(\boldsymbol{\theta}_n)\}^2}{2\sigma_p^2}\right], \tag{4}$$

$$\mathcal{L}^a_{m,n,t}\left(E^a_{m,n,t}(\boldsymbol{\theta}_n)\right) = \frac{1}{\sqrt{2\pi\sigma_a^2}} \exp\left[-\frac{\{E^a_{m,n,t}(\boldsymbol{\theta}_n)\}^2}{2\sigma_a^2}\right]. \tag{5}$$

$E^p_{m,n,t}(\boldsymbol{\theta}_n)$ and $E^a_{m,n,t}(\boldsymbol{\theta}_n)$ are the percentage and absolute error of observed and simulated mean velocity by $\boldsymbol{\theta}_n$ in the *m*-th road segment at time *t*, $\sigma_p$ and $\sigma_a$ are tolerance factors of the errors, respectively. Both the percentage and absolute error are used in the likelihood function of Equation (3) because the performance of the optimal model parameters estimation increases independent of growth stages of congestion owing to the sensitivity difference of the errors [18].

In class 1 and 2 road segments, the likelihood is determined as the weighted mean by the prior probability distribution of mean velocity with missing values $f(u)$,

$$\mathcal{L}'_{m,n,t}(\boldsymbol{\theta}_n) = \int_{u_{\min}}^{u_{\max}} f(u)\mathcal{L}^p_{m,n,t}\left(E^{p,\mathrm{mv}}_{m,n,t}(u,\boldsymbol{\theta}_n)\right) du \times \int_{u_{\min}}^{u_{\max}} f(u)\mathcal{L}^a_{m,n,t}\left(E^{a,\mathrm{mv}}_{m,n,t}(u,\boldsymbol{\theta}_n)\right) du. \tag{6}$$

$E^{p,\mathrm{mv}}_{m,n,t}(u,\boldsymbol{\theta}_n)$ and $E^{a,\mathrm{mv}}_{m,n,t}(u,\boldsymbol{\theta}_n)$ are the percentage and absolute error between $u$ and the simulated mean velocity by $\boldsymbol{\theta}_n$ in the *m*-th road segment at time *t*. The tolerance factors $\sigma_p$ and $\sigma_a$ are the same as in Equation (4) and (5). This paper adopts $\sigma_p = 20\%$ and $\sigma_a = 10$ km/h. From $\mathcal{L}'_{m,n,t}(\boldsymbol{\theta}_n)$, the likelihood in all road segments $\mathcal{L}_{n,t}(\boldsymbol{\theta}_n)$ is derived,

$$\mathcal{L}_{n,t}(\boldsymbol{\theta}_n) = \prod_m \mathcal{L}'_{m,n,t}(\boldsymbol{\theta}_n). \tag{7}$$

Then the weight $w_{n,t}(\boldsymbol{\theta}_n)$, which represents the similarity of the mean velocities to the simulation result by $\boldsymbol{\theta}_n$ at time *t* in all road segments is derived,

$$w_{n,t}(\boldsymbol{\theta}_n) = \left[\ln \mathcal{L}_{n,t}(\boldsymbol{\theta}_n)\right]^{-2}. \tag{8}$$

The inverse square of logarithmic likelihood returns positive values, which makes it easy to use in the particle filter process. After $w_{n,t}(\boldsymbol{\theta}_n)$ is derived for all simulation scenarios, it is normalized,

$$w_{n,t}(\boldsymbol{\theta}_n) \leftarrow \frac{w_{n,t}(\boldsymbol{\theta}_n)}{\sum_n w_{n,t}(\boldsymbol{\theta}_n)}. \tag{9}$$

Based on $w_{n,t}(\boldsymbol{\theta}_n)$, the particles set on $\boldsymbol{\theta}_n$ in the model parameter space are resampled. When $w_{n,t}(\boldsymbol{\theta}_n)$ is high, more particles are resampled. As the processes above are iterated at each time step, the number of particles at the model parameter set that produces a similar time trend of the observed result gradually increases. As a result, the posterior probability distributions of the model parameters are obtained from the distribution of the particles in step 4. For instance, in the case of Figure 3, the particle filter processes begin at time *t* = 50 minutes and iterated until *t* = 80 minutes. Then the posterior probability distributions of the four model parameters at *t* = 80 minutes are obtained. This paper regards the best scenario as the maximum a posteriori (MAP) model parameter set. In step 5, each missing value segment is replaced with the simulated mean velocity at the same time and position based on the best simulation scenario at the MAP model parameter set. These processes above are carried out every minute and the MAP model parameters are updated in real-time.

## 3. Traffic Data for Validation & Simulation

Our proposed method is validated on congestion data measured on two expressways in Japan managed by Central Nippon Expressway Company (C-NEXCO) as shown in Figure 2. The first congestion scenario was observed on Ken-O Expressway, or Metropolitan Inter-City Expressway, which is the outer-most circular expressway around Tokyo. A two-lanes section of the length 10 km including congestion with missing values is considered for the validation. The missing values account for 30–50% in the section. Congestion occurred due to a bottleneck around 8.5–10 km. After $t = 145$ minutes, the strength of this bottleneck was alleviated, and as a result, almost no missing values occured after $t = 180$ minutes. Hence, this paper considers congestion only before the time for evaluation of our proposed method. The available traffic counters in the 10 km validation section are located at 2.27, 3.86, 5.89, and 9.63 km. Among these, the traffic counter at 5.89 km is used to validate the accuracy of the missing value imputation because the total duration with the missing values is the longest. The other traffic counters are used to set the prior probability distributions in missing value segments using Equation (1).

The other congestion scenario was observed on Tomei Expressway, which links Tokyo and Nagoya and is a main expressway in Japan. A three-lanes section of the length 14 km including congestion with frequent missing values is considered for the validation. During congestion, approximately 80% of road segments are missing values in the 14 km section. Congestion occurred for $t = 35$–$325$ minutes due to the construction activity in the shoulder from 10.7 to 12.2 km. The available traffic counters in the 14 km validation section are located at 1.8, 3.46, 5.66, 7.86, and 10.84 km. The traffic counter at 7.86 km is used to validate the accuracy of the missing value imputation and others are used to set the prior probability distribution in Equation (1).

Table 1 lists the configurations of the cell-automata model for the traffic simulation. The spatial and time resolution follows a primary study of a theoretical model [19]. The total number of cells is $1000 \times 2$ and $1400 \times 3$ (road length $\times$ lane number) regarding the validation length and lane number on Ken-O and Tomei Expressway, respectively. The fast lane is set as the rightmost lane in the travel direction. Considering the actual vehicle speed, the speed limit on the fast lanes is set as 120 and 100 km/h on the Tomei and Ken-O Expressway models, respectively. The remaining lanes, i.e., slow lanes, are set with a speed limit of 100 and 80 km/h for each expressway model. The inflow rate of traffic volume at the origin of the expressway models is based on traffic counter data located at the 0 km position. For instance, the observed number of vehicles and instant velocity by the traffic counter is input to the simulation as the boundary condition at the origin. Table 2 summarizes the model parameter sampling for simulation scenario sets. These sampling ranges are determined by test simulations in advance with each expressway model. A total of 7350 ($15 \times 7 \times 7 \times 10$) scenarios are simulated with these model parameter sets.

Table 1. Cell-automata parameters for the simulation.

| Configuration | Value |
| --- | --- |
| Cell length (spatial resolution) | 10 meters |
| Time step size (time resolution) | 1.8 seconds |
| Corresponding velocity resolution | 20 km/h |
| Bottleneck position (Ken-O Expressway) | 8.6–9.8 km |
| Bottleneck position (Tomei Expressway) | 10.7–12.2 km |

Table 2. S-NFS model parameter sampling.

| Model parameter | Lower limit | Upper limit | Increment | Number of samplings |
| --- | --- | --- | --- | --- |
| Bottleneck random braking probability $p^{BN}$ | 0.26 | 0.54 | 0.02 | 15 |
| Random braking probability $p$ | 0.06 | 0.24 | 0.03 | 7 |
| Slow-to-start probability $q$ | 0.06 | 0.24 | 0.03 | 7 |
| Anticipation probability $r$ | 0.90 | 0.99 | 0.01 | 10 |

The lane-changing behavior is considered in the traffic simulation. In each time step, each car checks whether it can drive faster if it changes the lane by considering the velocity updating process of the S-NFS model and the surrounding vehicles. If the achievable velocity is higher in the adjacent lane, the vehicle moves to the lane with a certain probability. This paper adopts the lane-change probability of 10%, which is equivalent to a lane-changing vehicle at least once with the probability of 96.9% during a minute if all conditions for the lane changing are satisfied.

## 4. Results & Discussion

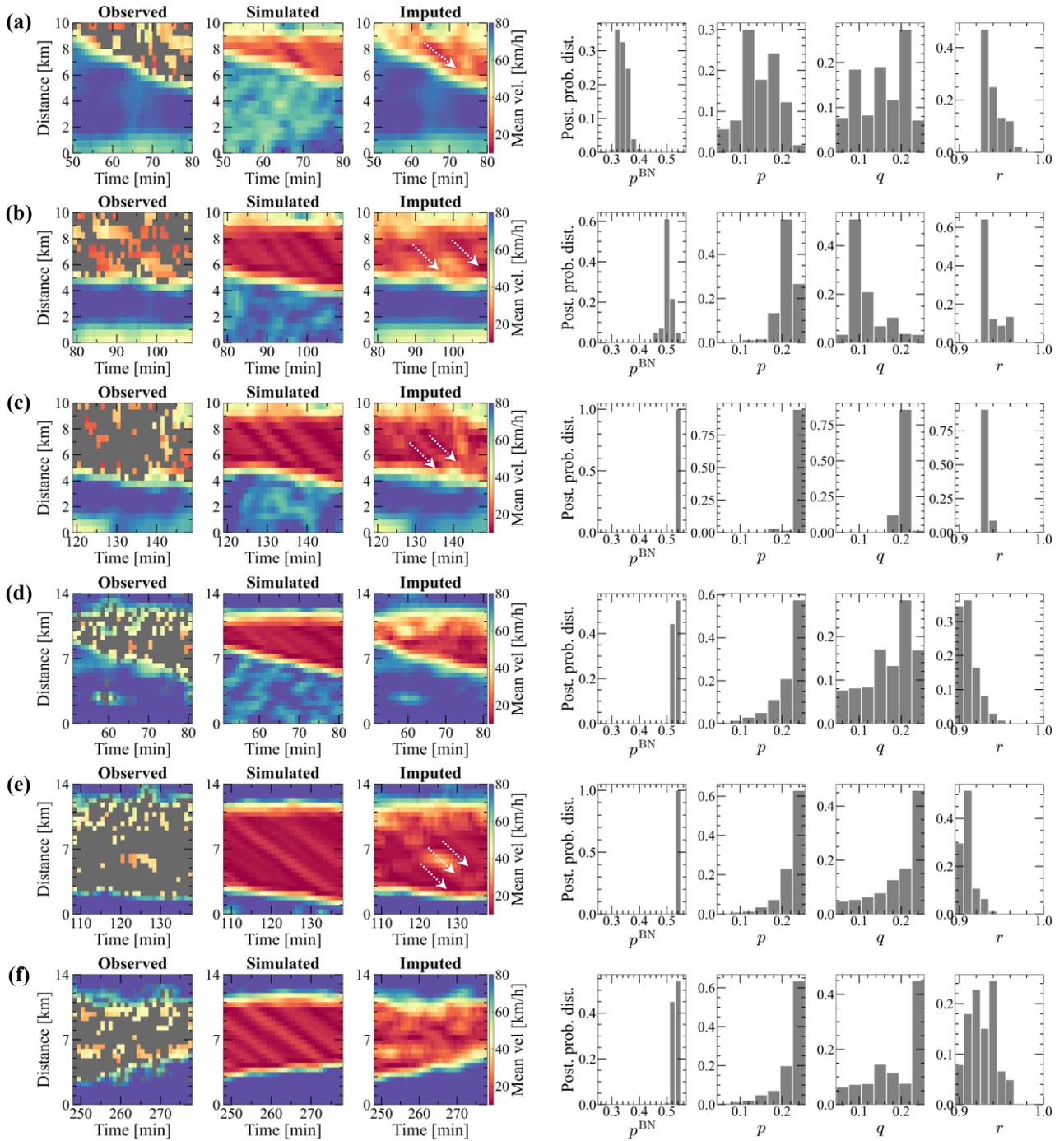

**Figure 6.** (From left to right) Mean velocities in observed data by DFOS, simulation result at the MAP model parameter set, and imputed results, and the posterior probability distributions of the model parameters on (a)–(c) Ken-O and (d)–(f) Tomei Expressway at the early, peak, and dissipating stage of congestion. White arrows in imputation results indicate density wave features.

Figure 6 shows heatmaps of observed, simulated, and imputed mean velocities at specific times on Ken-O and Tomei Expressway with posterior probability distributions of the model parameters. The simulation can reproduce the outline of congestion observed in mean velocity data. Moreover, velocity structures during congestion in missing value segments are reconstructed. For example, the imputation results of Figure 6 (a), (b), (c), and (e) show the density wave features highlighted by white arrows, which is a unique phenomenon to traffic flow.

To validate the imputed mean velocities in DFOS data, the deviation from the reference traffic counter data is examined. Note that the definition of mean velocity derived from DFOS and traffic counters differs. The former is based on trajectories in 500-m, 1-minute patches, whereas the latter measures instant velocity at the traffic counter location. Therefore, it is natural that these velocities are different, especially during congestion. Figure 7 shows the time-series data of observed and estimated mean velocity at missing value segments in DFOS data and traffic counter measurement results on Ken-O and Tomei Expressway

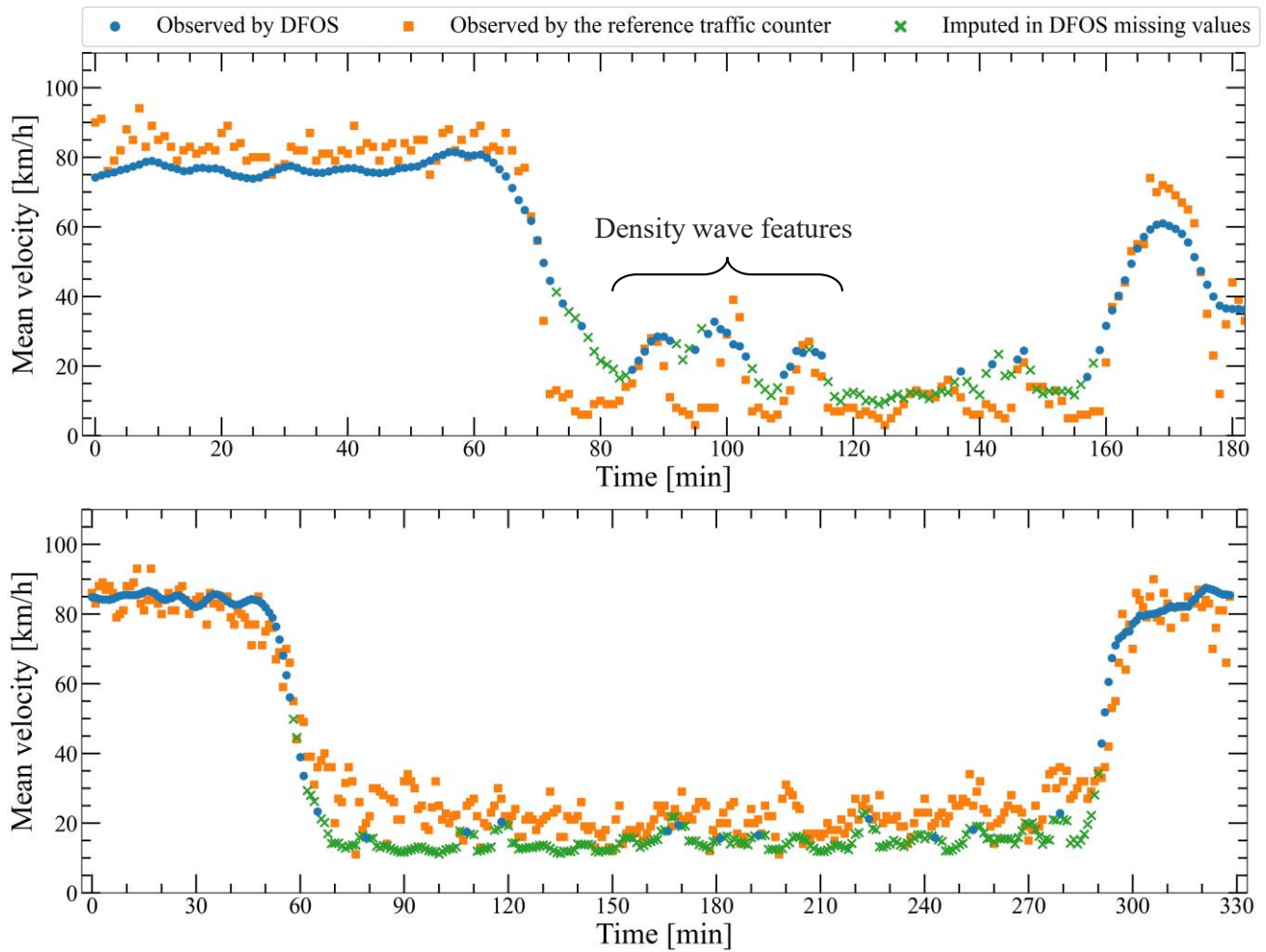

**Figure 7. Time series data of mean velocity obtained from DFOS, the traffic counter, and estimated mean velocity at missing value segments on (top) Ken-O Expressway and (bottom) Tomei Expressway.**

Figure 7 indicates that the deviation of mean velocity converges as congestion grows. In the early stage of congestion on Ken-O Expressway ($t \sim 75$–$80$ minutes), the imputed mean velocities are overestimated, i.e., the simulation reproduces more moderate congestion compared with the actual level. This is because the derived MAP model parameters do not correspond to a severe congestion scenario. During the early stage, the congestion section is insufficient to affect the particle filter result. Hence, the simulation result at the MAP model parameters underestimates the congestion length. In contrast to the results on Ken-O Expressway, the same tendency is not observed for Tomei Expressway at $t \sim 60$ minutes. A plausible interpretation for this discrepancy is due to the growth rate of the congestion length. For example, congestion on Ken-O Expressway grows relatively slowly (up to 5 km in 30 minutes), whereas that on Tomei Expressway grows rapidly (up to 10 km in 45 minutes). Thus, even in the early stage of congestion, the proportion of the congestion section used for the particle filter in the Tomei Expressway scenario is higher than that in the Ken-O Expressway scenario. As a result, the derived MAP model parameters reproduce a congestion length similar to the observation. Therefore, the deviations in imputed mean velocities in the Tomei Expressway scenario remain low. After the congestion length is almost steady, the deviation between imputed mean velocities and those obtained from the traffic counter is also low on both expressways, i.e., the deviation within 10 km/h.

Moreover, as denoted by "density wave features" in the top panel of Figure 7, the mean velocity fluctuations observed for $t = 85$–$115$ minutes are completely reconstructed from imputed mean velocities. The local maxima and minima in DFOS mean velocity data are in accord with those of the traffic counter. These fluctuations are caused by the density waves. They are observed in Figure 6 (b) as stripe patterns elongated from the left top to the right bottom. In the observed data of Figure 6 (b), mean velocities are observed in only moderate-density sections and those in high-density sections are missing values. Our imputation method can fill the mean velocities in the high-density parts.

Table 3 summarizes the mean absolute error (MAE) of observed and imputed mean velocities in DFOS data compared with the reference traffic counter data. The MAEs of the imputed mean velocities by our proposed method are comparable to those in non-missing value segments. Even in the severe scenario on Tomei Expressway, where most of the DFOS mean velocities are missing values during congestion, the MAE in missing value segments only increases by 1.49 km/h compared with the MAE of the non-missing value segments. Similarly, for the Ken-O Expressway scenario, the MAE of the imputed mean velocities increases by 0.24 km/h. These results indicate that the proposed method can estimate traffic flow in missing value segments with the same accuracy as the observed data.

**Table 3. Mean absolute errors (MAEs) of mean velocities in missing and non-missing value segments of DFOS data compared with the reference traffic counter.**

| Expressway | MAE of imputed mean velocities in missing value segments | MAE of observed mean velocities by DFOS (in non-missing value segments) |
|---|---|---|
| Ken-O | 8.58 km/h | 8.34 km/h |
| Tomei | 8.43 km/h | 6.94 km/h |

## 5. Conclusions

This paper has been proposed a missing value imputation method for mean velocity data observed by DFOS. We introduce prior probability distributions of mean velocity at missing values based on prior information, inferences from the observation, and other sensor data. The main conclusions are listed below.
1). The proposed method can reconstruct realistic traffic flow including density wave features in the imputed mean velocity data.
2). Comparing with the reference traffic counter data, the accuracy of the imputed mean velocities in missing value segments is comparable to the observed mean velocities in non-missing value segments. The MAE of imputed mean velocities increases only by at most 1.5 km/h compared with the non-missing values.

The proposed method overcomes the restriction of DFOS in practical cases and enhances its visibility in various traffic states. This can improve the wide-range applicability of DFOS for real-time traffic monitoring by visualizing traffic flow in dead zones of conventional traffic sensors.